\documentclass[twocolumn,showpacs,preprintnumbers,amsmath,amsfonts,amssymb,prb]{revtex4}
\preprint{%to be submitted to {\em Physical Review Lett}
}

\usepackage{graphicx}% Include figure files
\usepackage{dcolumn}% Align table columns on decimal point
\usepackage{bm}% bold math

\usepackage{color}
\definecolor{red}{rgb}{1,0,0}
\definecolor{green}{rgb}{0,1,0}
\definecolor{blue}{rgb}{0,0,1}

\begin{document}

\title{Structural evolution of bismuth sodium titanate induced by A-site non-stoichiometry: Neutron powder diffraction studies}

\author{I.-K. Jeong}
\altaffiliation {To whom correspondence should be addressed.
E-mail:Jeong@pusan.ac.kr}
\affiliation {Department of Physics
Education \& Research Center for
Dielectrics and Advanced Matter Physics, Pusan National University, Busan, 609-735, Korea.}

\author{Y. S. Sung}
\affiliation{Department of Material Science and Engineering,
POSTECH, Pohang, Gyeongbuk, 790-784, Korea.}
\author{T. K. Song}
\author{M.-H. Kim}
\affiliation{School of Nano and Advanced Materials Engineering,
Changwon National University, Gyeongnam, 641-773, Korea.}
\author{A. Llobet}
\affiliation{Lujan Neutron Scattering Center, Los Alamos National Laboratory, Los Alamos, NM 87545,
USA.}

%\email{jeong@lanl.gov}

\date{\today}% It is always \today, today,
             %  but any date may be explicitly specified

\begin{abstract}
We performed neutron powder diffraction measurements on (Bi$_{0.5}$Na$_{0.5+x}$)TiO$_3$ and (Bi$_{0.5+y}$Na$_{0.5}$)TiO$_3$ to study structural evolution induced by the non-stoichiometry.
Despite the non-stoichiometry, the local structure ({\it r}~$\leq$~3.5~{\AA}) from the pair distribution function analysis is barely affected by the sodium deficit of up to -5~mol\%. With increasing pair distance, however, the atomic pair correlations weaken due to the disorder caused by the sodium deficiency.
Although the sodium and the bismuth share the same crystallographic site, their non-stoichiometry have rather opposite effects as revealed from a distinctive distortion of the Bragg peaks.
In addition, Rietveld refinement demonstrates that the octahedral tilting is continually suppressed by the sodium deficit of up to -5~mol\%. This is contrary to the effect of the bismuth deficiency, which enhances the octahedral tilting.
\end{abstract}

\maketitle

\section{Introduction}

(Bi$_{0.5}$Na$_{0.5}$)TiO$_3$ (BNT) based solid solutions are one of
the most promising candidates for replacing lead-containing electromechanical ceramics such as Pb(Zr$_{1-x}$Ti$_x$)O$_3$, and have been extensively studied~\cite{ma;afm13, simons;apl11, Jones;sensors10, ranjan;apl09, Rodel;jacs09}.
On the substitution of the sodium by the potassium ion,
BNT undergoes a phase transition
from rhombohedral to tetragonal structure~\cite{Jones;pd02}.
Similarly, solid solution of BNT with BaTiO$_3$ exhibits rich structural phases~\cite{simons;apl11, ma;prl12} as a function of composition and poling field with enhanced ferroelectric response near the morphotropic phase boundary~\cite{Rodel;jacs09}.
In addition, recent works report that a doping on perovskite A- and B-site~\cite{watanabe;f07, davies;jacs11,aksel;prb12} modifies physical properties such as depolarization temperature, $T_d$ and piezoelectric constant ($d_{33}$).
For example, a doping of lanthanide ions~\cite{watanabe;f07, aksel;jap12} on the A-site
decreases the depolarization temperature, $T_d$. Contrary, the B-site substitution of the iron and the manganese~\cite{davies;jacs11,aksel;prb12} enhances the $T_d$.

\begin{figure}[b]\hspace{-1.0cm}
\includegraphics[angle=0,scale=0.65]{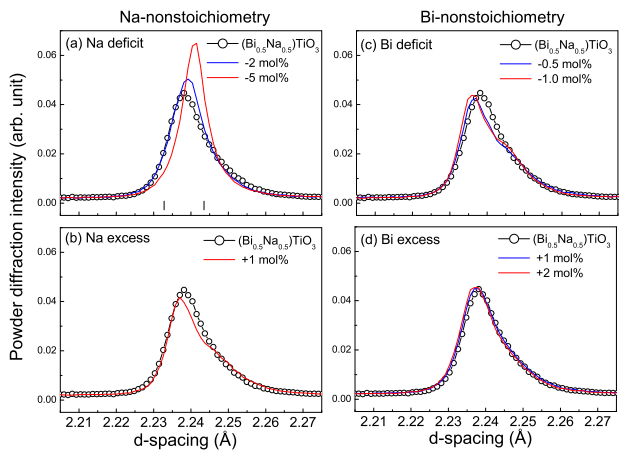}
\caption { (color online) Effects of non-stoichiometry on (202) and (006) Bragg peaks. Symbol (-o-) represents the Bragg peak of stoichiometric (Bi$_{0.5}$Na$_{0.5}$)TiO$_3$ and solid lines are those of non-stoichiometric samples. The tick marks in (a) are Bragg peak positions for the stoichiometric sample. Note that the non-stoichiometry of sodium (a,b) and bismuth (c,d) have distinct influences on Bragg peak positions and shape.}
\label{fig;fig1}
\end{figure}

On the crystal structure of a doped BNT,
it is found that the crystal structure approaches cubic phase with increasing addition of La~\cite{herabut;jacs97}. Also, a slight increase of the cell volume by the Fe doping is observed~\cite{aksel;prb12}. These concurrent structural modifications with the evolution of the $T_d$ hint an underlying role of the structure on the depolarization temperature.
In this paper, we report a structural evolution of non-stoichiometric (Bi$_{0.5}$Na$_{0.5+x})$TiO$_3$ and (Bi$_{0.5+y}$Na$_{0.5}$)TiO$_3$ from neutron powder diffraction measurements.
Using both atomic pair distribution function analysis~\cite{egami;bk12} and Rietveld refinement we investigate the local structural distortion below the atomic pair distance {\it r}$\sim$3.5~{\AA} as well as the long-range octahedral tilting induced by the sodium and the bismuth non-stoichiometry, respectively.

\section{Experiments}

Stoichiometric (Bi$_{0.5}$Na$_{0.5}$)TiO$_3$, non-stoichiometric (Bi$_{0.5}$Na$_{0.5+x}$)TiO$_3$ ($x$=-5, -2, +1 mol\%) and (Bi$_{0.5+y}$Na$_{0.5}$)TiO$_3$ ($y$=-1, -0.5, +1 mol\%) ceramic samples were prepared using a solid-state reaction. After calcination and intermediate ball milling powders were pelletized and sintered at 1150 $^\circ$C for 2 h in air. The apparent densities of the pellets after the sintering were above $\sim$ 95 \% of the theoretical values, indicating that all samples were prepared consistently~\cite{sung;apl10, sung;apl11}.
Ceramic samples were crushed into fine powders and then annealed to relieve strain.
Time-of-flight neutron powder diffraction measurements were performed on the NPDF instrument at the Los Alamos Neutron Science Center. Powder samples were loaded in vanadium cans and then mounted in a closed-cycle helium cryostat with Helium exchange gas.
All measurements were performed at $T$=180~K to reduce thermal contribution to structural features.

\begin{figure}[b]\hspace{-1.0cm}
\includegraphics[angle=0,scale=0.8]{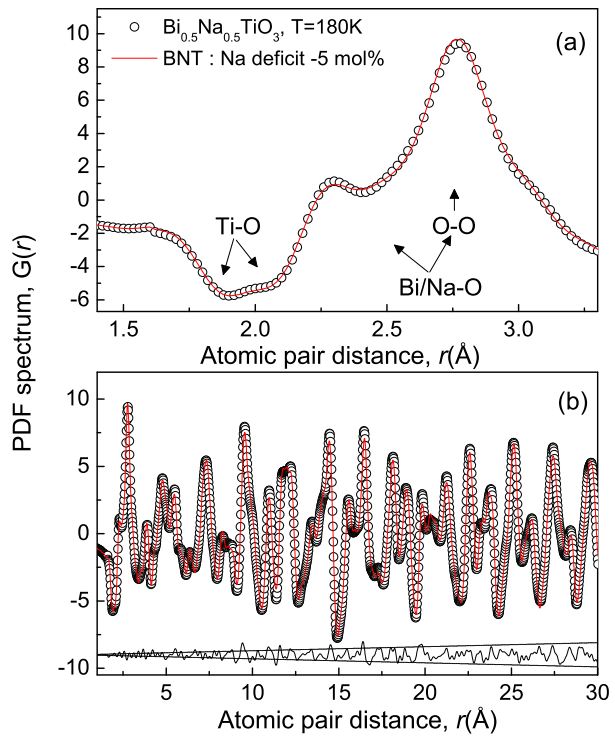}
\caption { (color online) Neutron PDF spectrum of stoichiometric (Bi$_{0.5}$Na$_{0.5}$)TiO$_3$ (open circle) and sodium -5~mol\% deficit BNT (solid line) at T=180~K obtained via Fourier transform of structure function with Q$_{\rm max}$=25~{\AA}$^{-1}$. (a) The first few peaks correspond to Ti-O, Bi/Na-O, and O-O bond. (b) The difference curve at the bottom indicates that the discrepancy between the two spectra increases with increasing pair distance due to  the sodium deficiency.}
\label{fig;fig2}
\end{figure}

\section{Results and Discussion}

Figure~1 summarizes the effects of the non-stoichiometry on (202) and (006) Bragg peaks. The tick marks in (a) are peak positions for the stoichiometric BNT represented by the symbol (-o-). Solid lines are those of the non-stoichiometric samples. First, note that the non-stoichiometry of the sodium (a,b) and the bismuth (c,d) have rather opposite effects on the Bragg peak positions and shape.
For the sodium deficient samples, the Bragg peak shifts toward higher d-spacing. In addition, the peak becomes sharper and symmetric with increasing deficiency, much like the case of the lanthanum doped BNT~\cite{herabut;jacs97}.
In contrast, the Bragg peak position shifts toward lower d-spacing and the shoulder around d-spacing 2.25 {\AA} becomes pronounced due to the bismuth deficiency. Unlike the sodium excess, the bismuth excess barely affects the Bragg peak positions and shape.
Overall, these results emphasize the distinctive role of the sodium and the bismuth deficiency/excess on the lattice distortion. %

The A-site deficiency of Bi/Na ion generates oxygen vacancies~\cite{sung;apl10}
which induce a distortion of the octahedral unit, and thus affects a local displacement of Ti ion~\cite{jeong;prb11} and induces a leakage conductivity in ferroelectric BNT~\cite{li;nm14}.
To examine the local structural distortion induced by the A-site deficiency we performed neutron pair distribution function (PDF) analysis on stoichiometric BNT and sodium -5~mol\% deficit BNT.
PDF spectra were obtained via Fourier transform of experimental total scattering structure functions~\cite{egami;bk12} with Q$_{\rm max}$=25~{\AA}$^{-1}$ after data reduction using the program PDFgetN~\cite{pdfgetn}.
In Fig.~2(a), the first few peaks correspond to Ti-O, Bi/Na-O, and O-O bonds.
The Ti-O bond appears as a negative peak due to a negative neutron scattering length of the titanium. In addition, the doublet peak shape indicates that the Ti ion is off-centered in an oxygen octahedron~\cite{jeong;zk11, keeble;afm13}.
Note that the PDF spectra of the stoichiometric (open circles) and sodium -5~mol\% deficit (solid line) BNT overlap quite nicely up to the pair distance {\it r}$\sim$3.5 {\AA}. This result indicates that the local structures around the Ti ion and of the oxygen octahedron are little affected by the sodium deficiency.
However, as the difference curve in Fig.~2(b) indicates, the deviation between the two spectra becomes larger at higher-$r$. The increasing discrepancy reflects a weakening of the atomic correlations due to a structural disorder in the sodium deficient sample.
\begin{figure}[b]\hspace{-1.0cm}
\includegraphics[angle=0,scale=0.8]{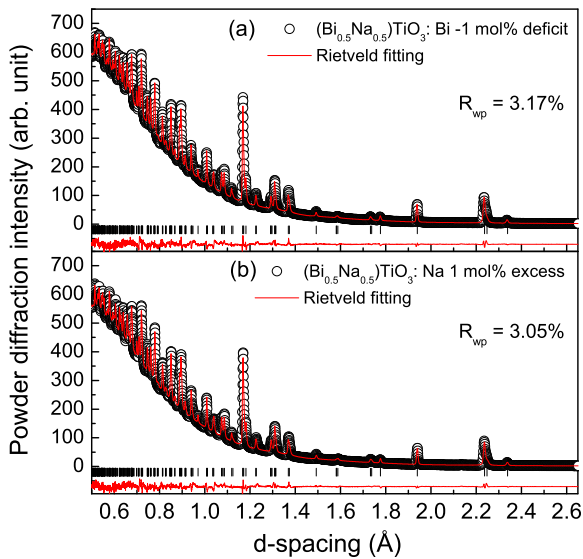}
\caption { (color online) Neutron powder diffraction patterns (open circle) of a bismuth -1~mol\% deficit and a sodium 1~mol\% excess sample, respectively. Solid line represents a Rietveld fitting using rhombohedral $R3c$ structure at 180~K. Tick marks represent Bragg peak positions. Also shown are the difference curves between experimental and model patterns.}
\label{fig;fig3}
\end{figure}
We now focus on the long-range structural evolution of BNT induced by the non-stoichiometry. The crystal structure of ferroelectric BNT is somewhat controversial. Originally, it was reported as rhombohedral $R3c$ structure by Jones and Thomas~\cite{Jones;acsb02} by using neutron powder diffraction measurements. In the subsequent high resolution x-ray diffraction measurements, however, a subtle peak splitting was observed in a sintered powder sample and monoclinic $Cc$ structure~\cite{aksel;apl11} was proposed. More complex picture was proposed by Rao {\it et al.} on electrically poled and thermally annealed specimens~\cite{rao;prb13}.
The authors found that both rhombohedral and monoclinic phases coexisted in the thermally annealed sample. In addition, the fraction of the $R3c$ phase was increased by applying an external electric field and a mechanical grinding.
In recent, the structural ambiguity of BNT has been reinvestigated by using a convergent beam electron diffraction of unpoled single crystals. The results show that defect-free BNT has an average $R3c$ symmetry over a few nanometer length scale. Near defects such as antiphase boundary and domain walls, however, the material will have a monoclinic symmetry~\cite{beanland;prb14}.
In our Rietveld analysis we tried both rhombohedral and monoclinic phases to refine neutron powder diffraction pattern of the stoichiometric BNT at 180~K. We found that the fitting with rhombohedral $R3c$ phase was more stable and easily converged with less number of parameters than that using monoclinic $Cc$ phase. Thus, we kept the $R3c$ phase and used it for the refinement of the other non-stoichiometric samples.

Figure~3 shows representative refinements on a bismuth deficit (-1~mol\%) and a sodium excess (1~mol\%) samples using GSAS~\cite{gsas} interfaced by EXPGUI~\cite{expgui}.
Solid line represents a Rietveld fitting using rhombohedral $R3c$ structure~\cite{Jones;acsb02}. Tick marks indicate Bragg peak positions. Also shown is a difference curve between the experimental and the model patterns.
In the Rietveld refinements, lattice and thermal parameters as well as atomic positions were refined along with absorption and background corrections. The bismuth and the sodium ions are assumed to share the same crystallographic position. High quality of the fitting confirms that all samples maintain the average $R3c$ structure.
Despite the presence of heavy ion (Bi), the atomic positions of the oxygen ion were reliably refined due to a large scattering contrast of oxygen in the neutron diffraction. Based on these structural information we calculated the octahedral tilting angle~\cite{Jones;acsb02} as a function of the sodium and the bismuth non-stoichiometry, respectively.

\begin{figure}[b]\hspace{-1.0cm}
\includegraphics[angle=0,scale=0.8]{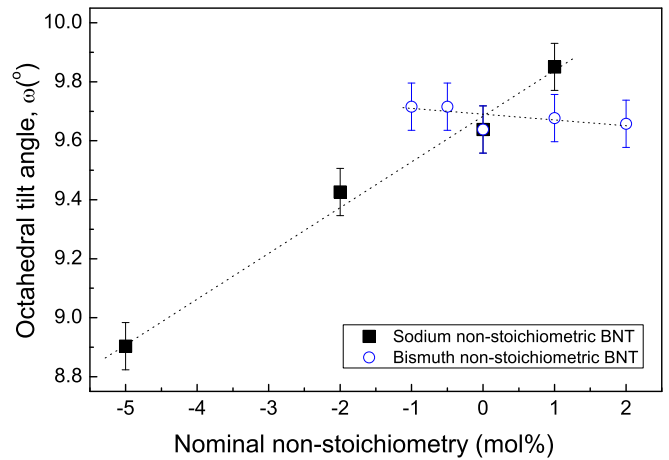}\hspace{-1.0cm}
\caption { (color online) (a) Antiphase octahedral tilting (a$^-$a$^-$a$^-$ Glazer tilt system) of non-stoichiometric BNT. Closed square and open circle are for sodium and bismuth non-stoichiometric samples, respectively. Dotted lines are guides to the eye. The inset shows the projection of the rhombohedral structure with a counter-clockwise tilting.}
\label{fig;fig4}
\end{figure}

In the rhombohedral BNT, the antiphase octahedral tilting (a$^-$a$^-$a$^-$ Glazer tilt system)~\cite{glazer;acb72,megaw;aca75} is coupled with cation displacements~\cite{Jones;acsb02} and plays an important role for the stability of the ferroelectric phase.
The tilting appears at around 670~K and increases with decreasing temperature, saturating to $\sim 9.5^\circ$ at 5~K~\cite{Jones;acsb02}.
Figure~\ref{fig;fig4} shows the octahedral tilting angle $\omega$ as a function of the sodium ($\blacksquare$) and the bismuth ($\bigcirc$) non-stoichiometry at 180~K.
For the sodium non-stoichiometric BNT ($\blacksquare$), the antiphase octahedral tilting angle
evolves with a positive slope as a function of the non-stoichiometry.
At the sodium deficit of -5~mol\%, the tilting angle is suppressed to $\sim$8.9$^\circ$. With the sodium excess of 1~mol\%, however, the tilting is enhanced to $\sim9.85^\circ$.
Effectively, the change of the tilting angle between -5~mol\% $\leq x \leq$~1~mol\% is equivalent~\cite{Jones;acsb02} to the the temperature variation of about 100~$^\circ \rm C$, and reflects the instability of the ferroelectric lattice to the sodium non-stoichiometry.
For the bismuth non-stoichiometric BNT ($\bigcirc$), the tilting trend is more or less opposite to that of the sodium non-stoichiometric BNT i.e. the tilting increases slightly with the deficiency of the bismuth, exhibiting a negative slope.
This is an interesting contrast, indicating that the sodium and the bismuth play a distinct role on the crystal structure of BNT. In fact, recent studies using neutron total scattering analysis show that the displacement of Bi$^{3+}$ is about twice larger~\cite{keeble;afm13} than that of Na$^+$.
As a result, an octahedron tilting which is coupled to the A-site ionic displacements would be different around the sodium and the bismuth ions, respectively.
The disparity of the octahedral tilting around Bi$^{3+}$ and Na$^+$ imposes an important structural feature such as nano-scale tilting disorder~\cite{levin;afm12}, which is observed by transmission electron microscopic studies.
With A-site non-stoichiometry, the tilting disorder will be increased due to a fluctuation of the local chemistry and the resultant variation of the octahedral tilting angle.
\section{Conclusion}

Using neutron powder diffraction studies,
we find that the local structure of BNT below the atomic pair distance $r\leq$~3.5 {\AA} is little influenced, contrary to the noticeable suppression/enhancement of the long-range oxygen octahedral tilting by the A-site non-stoichiometry. In addition, we report that
the sodium and the bismuth non-stoichiometry induce a rather opposite octahedral tilting evolution. These results provide a structural hint for a disparity of the tilting behavior around Bi$^{3+}$ and Na$^+$, which leads to a local tilting disorder.

\acknowledgments

We are grateful to K. Page and J. Siewenie for helping with the data collection.
This work was supported by the National Research Foundation of Korea grant funded by the Korean Government NRF-2013R1A1A2012499.
This work has benefited from the use of NPDF at the Lujan Center at Los Alamos Neutron Science Center, funded by DOE Office of Basic Energy Sciences. Los Alamos National Laboratory is operated by Los Alamos National Security LLC under DOE Contract DE-AC52-06NA25396. 

%\bibliography{d:/papers/bib/2010-2015,%
%d:/papers/bib/2001-2010,%
%d:/papers/bib/1994-2000,%
%d:/papers/bib/bnkt2,%
%d:/papers/bib/2008-pzn-pt,%
%}

%merlin.mbs apsrev4-1.bst 2010-07-25 4.21a (PWD, AO, DPC) hacked
%Control: key (0)
%Control: author (8) initials jnrlst
%Control: editor formatted (1) identically to author
%Control: production of article title (-1) disabled
%Control: page (0) single
%Control: year (1) truncated
%Control: production of eprint (0) enabled
%

\end{document}